\documentclass[11pt]{article}

\usepackage{graphicx}
\usepackage[utf8]{inputenc}
\usepackage{float}
\usepackage{amsmath,amssymb}

\title{Temperature Anisotropy  of the CMBR  and the  Non-zero Cosmological Constant}
\author{Jan Schee$^1$ and Evariste N. Boj$^2$\\
	\textit{Research Centre for Theoretical Physics and Astrophysics},\\ 
	\textit{Institute of Physics}, \\
	\textit{Silesian University in Opava},\\
	 \textit{Bezru\v{c}ovo n\'{a}m. 13},\\
	  \textit{CZ-746 01 Opava, Czech Republic}\\
	  email 1: jan.schee@physics.slu.cz, \\
	  email 2: evariste.boj@gmail.com}

\date{}
\newcommand{\diff}{\mathrm{d}}

\newenvironment{dedication}
  {\clearpage           
   \thispagestyle{empty}
   \vspace*{\stretch{1}}
   \itshape             
   \raggedleft          
  }
  {\par 
   \vspace{\stretch{3}} 
   \clearpage           
  }

\begin{document}
\maketitle
\begin{abstract}
	We analyze the effect of a spherically symmetric clump on the anisotropy of the  cosmic microwave background radiation temperature in the framework of the standard model of  cosmology with a non-zero cosmological constant $\Lambda$ and we show that it weakens the Rees-Sciama effect.  
\end{abstract}

\section*{Introduction}
The observations of the temperature anisotropies of the Cosmic Microwave Background Radiation (CMBR) , $\Delta T/T_0$  are a key source of  information about the nature of the distribution of cosmic matter. The level of $\Delta T/T_0\sim 10^{-5}$ is believed to correspond to small density fluctuations of the cosmic matter being at the origin of large cosmic structures. The satellite missions COBE, WMAP, Planck \cite{COBE:1992:ApJ,WMAP:2003:ApJ,Planck:2011:AA} delivered detailed informations about the temperature anisotropies of the CMBR, with level of precision exceeding $10^{-5}$ \cite{Bennett_etal:2013,Planck:2018:AA}.The temperature anisotropies of the CMBR are of two fundamental kinds (here we constraint ourselves only to the temperature anisotropies of the CMBR that are gravitationally induced). The primary anisotropies are associated with the energy density fluctuations of the cosmic matter at a cosmological redshift $\sim 1300$, i.e. during the  era of the "recombination". The associated rise of local gravitational potentials leads to the Sachs-Wolfe effect \cite{Sachs_Wolfe:1967:ApJ}, during which, the radiation climbs out of the potential well due to a local overdensity of matter  and thus modifies the cosmological redshift, the radiation then looks to arrive from different cosmic era and subsequently with associated  different effective temperatures than the radiation that propagates through the homogeneous Universe. The secondary temperature anisotropy is induced by the  evolving density fluctuations at  the era between the  "recombination" period and now, i.e. the era  that is characterised with a cosmological redshift $z\lesssim 10$. This region is formed by a gravitationally bound system (a large galaxy, a cluster of galaxy) or by a large void, that is separated from the expanding Universe by a vacuum region and its boundary expands together with the expansion rate of the  Universe. The CMBR photon propagating through this region suffers a net gravitational redshift due to the expanding boundary. This is the Rees-Sciama effect \cite{Rees-Sciama:1968:Nature} and it  is the key effect we focus on in this paper. The Rees-Sciama effect was considered in detail in many research paper, let us mention the key ones that directly influenced our research \cite{Dyer:1976:MNRAS,Mes-Mol:1996:ApJ, Stu-Schee:2006}. There the standard Einstein-Strauss \cite{Einstein-Straus:1945:RevModPhys,Stuchlik:1984:} vacuola model was used to describe the gravitationally bounded clusters immersed in the Friedman-Lemaitre-Robertson-Walker (FLRW) Universe.    

Recent observations of the  temperature anisotropies of the CMBR   indicate that the cosmological constant, $\Lambda >0$, plays a very important role in the evolution of the Universe. It is strongly believed that the Universe undergoes an accelerated expansion, now, which is driven by the dark energy, that is in the simplest scenario associated with a positive cosmological constant \cite{Bennett_etal:2013,Planck:2018:AA}.

Here, we analyse the effect of $\Lambda>0$ on the  temperature anisotropies  of the CMBR  caused by the  Rees-Sciama effect, using the  Einstein-Strauss model. We integrate the  exact equations of motions of the  photon through the inhomogeneity, which we call a clump \cite{Dyer:1976:MNRAS}. We present two kinds of clumps. One, modelled by the  Schwarzschild-de Sitter black hole spacetime and the second  modelled by a  constant density halo.

The paper is organised as follows, In Section 1 we discuss details of the clump model, Section 2 is devoted to the definitions and the calculations of the temperature anisotropies of the CMBR  caused by the  clump, in Section 3 we explain our simulation setups and present our results, in Section 4  we discuss and conclude the results  of our research.

\section{The Inhomogeneity Model}
\begin{figure}[H]
	\begin{center}
		\includegraphics[scale=0.5]{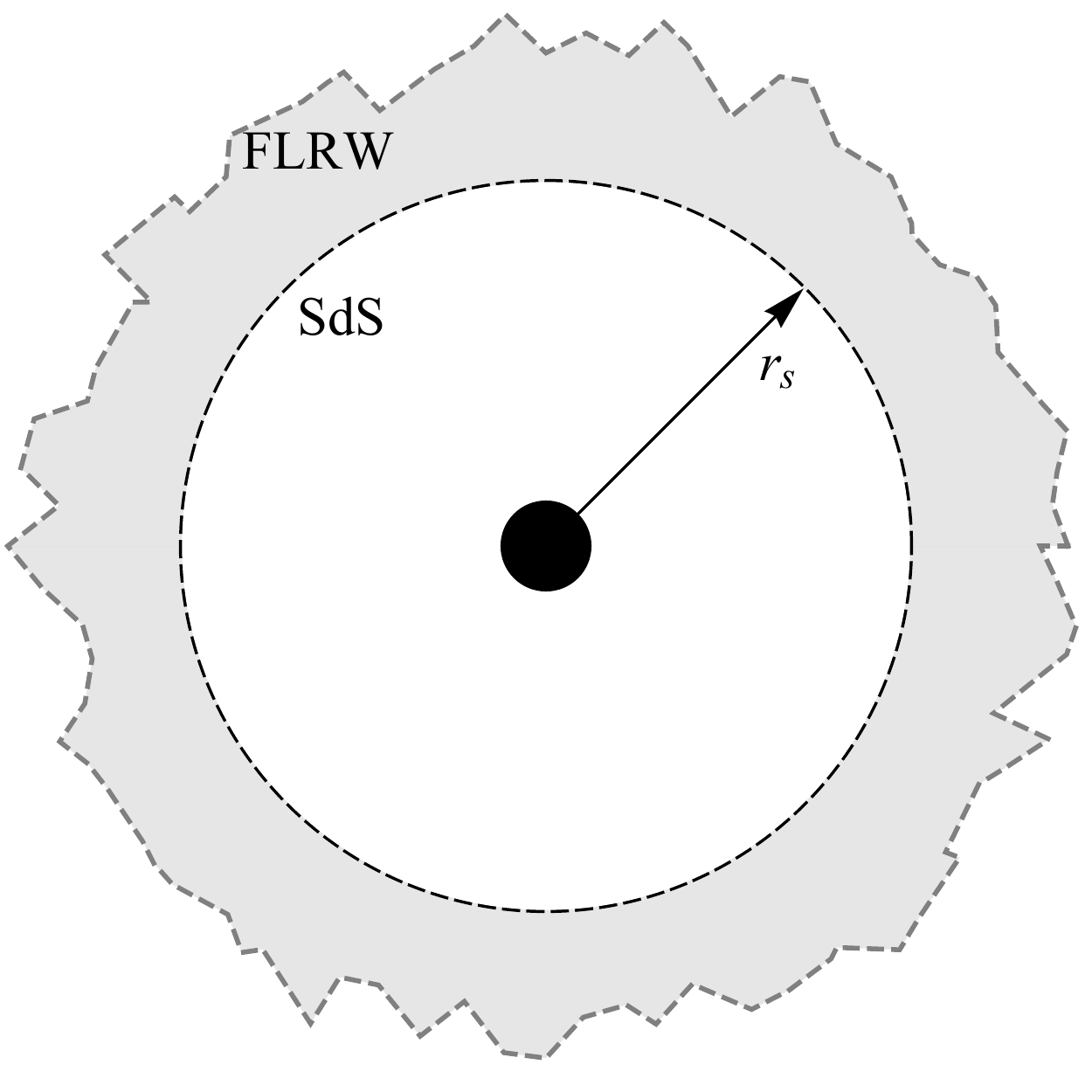}
		\caption{Scheme of the clump represented by a Schwarzschild-de Sitter black hole spacetime attached to the FLRW spacetime via the matching hypersurface.\label{Fig1}}
	\end{center}
\end{figure}

\begin{figure}[H]
	\begin{center}
		\includegraphics[scale=0.5]{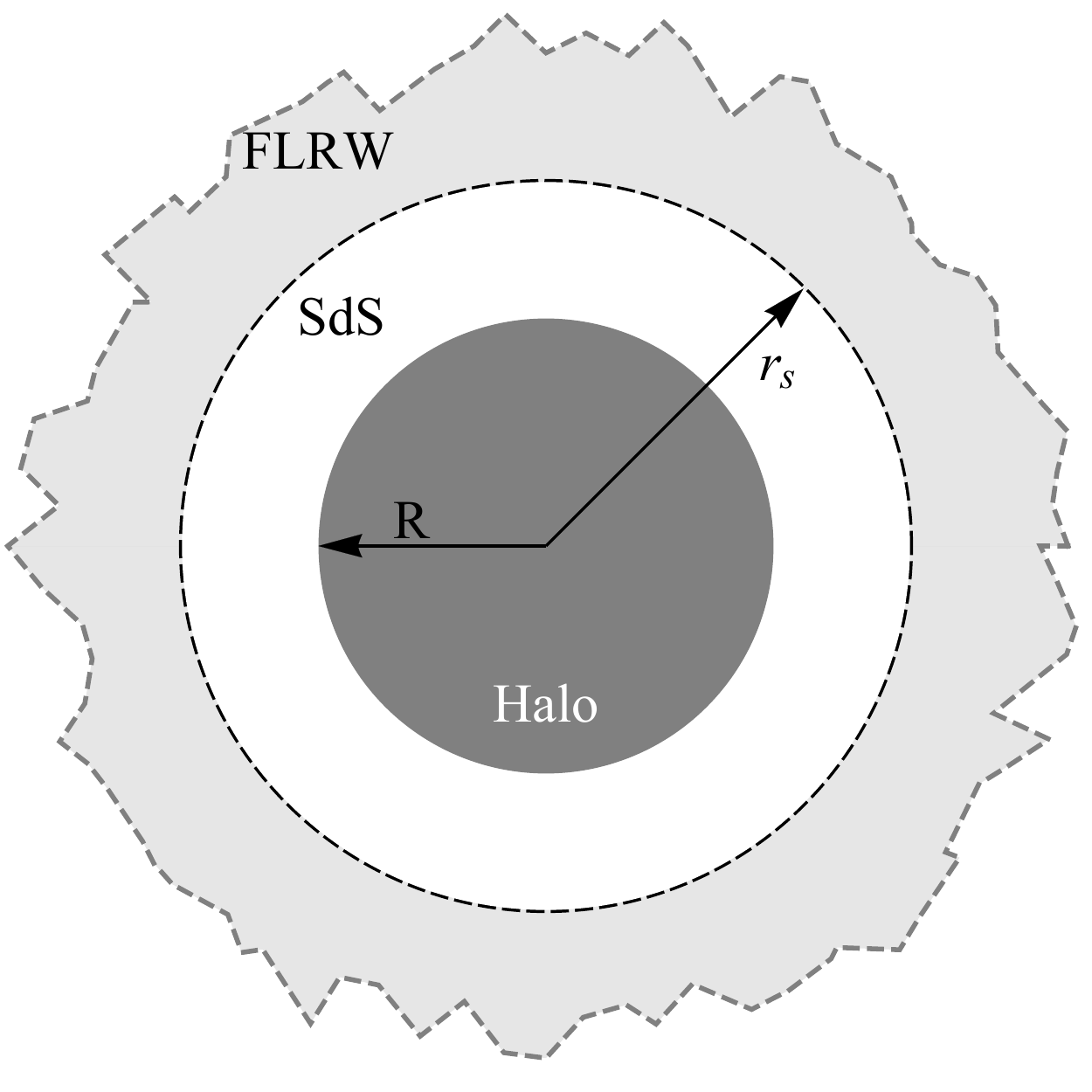}
		\caption{Scheme of the clump represented by a constant density halo spacetime with  a radius $R$. The vacuum Schwarzschild-de Sitter region  is attached to the FLRW spacetime via the matching hypersurface.\label{Fig2}}
	\end{center}
\end{figure}

The inhomogeneity (the Clump) is represented by the Einstein-Strauss family model \cite{Einstein-Straus:1945:RevModPhys,Stuchlik:1984:}. The idea is to construct a model of the universe containing an inhomogeneity to analyse  its effect on the CMBR effective temperature. Here we consider a very simple model of the homogenous and isotropic FLRW universe equipped with the coordinates $(\tau_F,\chi,\theta,\phi)$, where we replace a region of a constant commoving radius $\chi_s$ with a spherically symmetric Schwarzschild-de Sitter (SdS) spacetime or with the clump formed by a perfect fluid halo with constant density  with an external SdS vacuum region. In both cases, the SdS region is matched to the FLRW region through the  matching hypersurface (see Figs \ref{Fig1} and \ref{Fig2}). This approach was inspired by papers \cite{Dyer:1976:MNRAS,Mes-Mol:1996:ApJ,Stu-Schee:2006}.

\subsection*{The FLRW Spacetime}
The spacetime interval of the FLRW in the  commoving isotropic coordinates $(\tau_F,\chi,\theta,\phi)$ reads
\begin{equation}
	\diff s^2 = -\diff \tau_F^2 + a^2(\tau_F)\left[\diff\chi^2 + \Sigma_k^2(\chi)\left(\diff\theta^2 + \sin^2\theta\diff\phi^2\right)\right]
\end{equation}
where is
\begin{equation}
	\Sigma_k(\chi)=\left\{
		\begin{array}{cc}
			\sin\chi & \textrm{for}\quad k=1,\\
			\chi & \textrm{for}\quad k=0,\\
			\sinh\chi & \textrm{for}\quad k=-1,
		\end{array}
	\right.
\end{equation}
and $\tau_F$ is the proper time of fundamental cosmic observers comoving with the Universe, $\chi$ is the comoving radial coordinate, and $\theta$ and $\phi$ are usual latitudinal and azimuthal coordinates on the sphere.
The scale parameter $a(\tau_F)$ is determined by the Friedmann equation 
\begin{equation}
	H^2 = \frac{8\pi\,G}{3}\rho + \frac{\Lambda}{3} - \frac{k}{a^2}\label{Friedmann}
\end{equation}
where is $H\equiv\dot{a}/a$ and $\dot{}$-operator is just the derivative with respect to the  cosmic time $\tau_F$.

\subsection*{The Schwarzschild-de Sitter black hole}
The static, spherically symmetric solution of the vacuum Einstein equations with a non-zero cosmological constant $\Lambda$ is the SdS spacetime. In the usual Schwarzschild coordinates $(t,r,\theta,\phi)$ it reads
\begin{equation}
	\diff s^2=-f(r)\diff t^2+\frac{1}{f(r)}\diff r^2 + r^2\left(\diff\theta^2 + \sin^2\theta \diff\phi^2\right)
\end{equation}
where is
\begin{equation}
	f(r)=1-\frac{2M}{r}-\frac{\Lambda}{3}r^2.
\end{equation}
The spacetime posses two horizons, which are  both solutions of the equation $f(r)=0$, the black hole horizon $r_h$ and the cosmological horizon $r_c$ given by the formulae
\begin{eqnarray}
	r_h &=& \frac{2}{\sqrt{\Lambda}}\cos\left(\alpha - \frac{2\pi}{3}\right),\\
	r_c &=& \frac{2}{\sqrt{\Lambda}}\cos\left(\alpha\right)
\end{eqnarray}
where is 
\begin{equation}
	\alpha\equiv \frac{1}{3} \arccos\left(-3M\sqrt{\Lambda}\right).
\end{equation}
There is another important radius here, the static radius $r_s$, where the repulsion caused by a  positive cosmological constant is balanced by the  gravity  of the central body and is located at
\begin{equation}
	r_s = \left(\frac{3M}{\Lambda}\right)^{1/3}.
\end{equation}

\subsection*{The constant density perfect fluid halo}
Solving the Tolman-Openheimer-Volkoff (TOV) equation with non-zero cosmological constant for a  perfect fluid with uniform density, one arrives to the spacetime interval in the form
\begin{equation}
	\diff s^2=-f(r)\diff t^2 + h(r)\diff r^2 + r^2\left(\diff \theta^2 + \sin^2\theta\diff\phi^2\right)
\end{equation}
where the metric functions read
\begin{equation}
	f(r)=\left\{
		\begin{array}{cc}
			\left[\frac{3}{2}\sqrt{1-\frac{2M}{R}-\frac{\Lambda}{3}r^2}-\frac{1}{2}\sqrt{1-\left(\frac{2M}{R^3}+\frac{\Lambda}{3}\right)r^2}\,\right]^2 & \textrm{for}\quad r<R\\
			1-\frac{2M}{r}-\frac{\Lambda}{3}r^2 & \textrm{for}\quad r\geq R
		\end{array}
	\right.
\end{equation}
and
\begin{equation}
	h(r)=\left\{
		\begin{array}{cc}
			\left[1-\frac{2M}{r}\left(\frac{r}{R}\right)^3-\frac{\Lambda}{3}r^2\right]^{-1} & \textrm{for}\quad r<R,\\
			\left[1 - \frac{2M}{r} -\frac{\Lambda}{3}r^2\right]^{-1}
		\end{array}
	\right.
\end{equation}
where $R$ is the radius of the halo. Of course, in this spacetime there is only the cosmological horizon, $r_c$, the potential presence of  the static radius, $r_s$,  depends on the size of the clump $r_S$. However, the present value of the cosmological constant is very small, $\Lambda\sim 10^{-52}\mathrm{m}^{-2}$. Considering the mass of a  typical cluster of galaxies  $M\sim 10^{18}\mathrm{M}_{\odot}$, the static radius will be at $r_s\approx 250\mathrm{Mpc}$ and the cosmological horizon at $r_c\approx 5000\mathrm{Mpc}$. In the era of interest, the static radius can be within the clump, while the cosmological horizon is well outside the clump.  There are no black hole horizon.

\subsection*{The Matching hypersurface}
The matching hypersurface is generated by the fundamental observers at the FLRW side and by the radially receding observers at the SdS side. 
\begin{figure}[H]
	\begin{center}
		\includegraphics[scale=0.4]{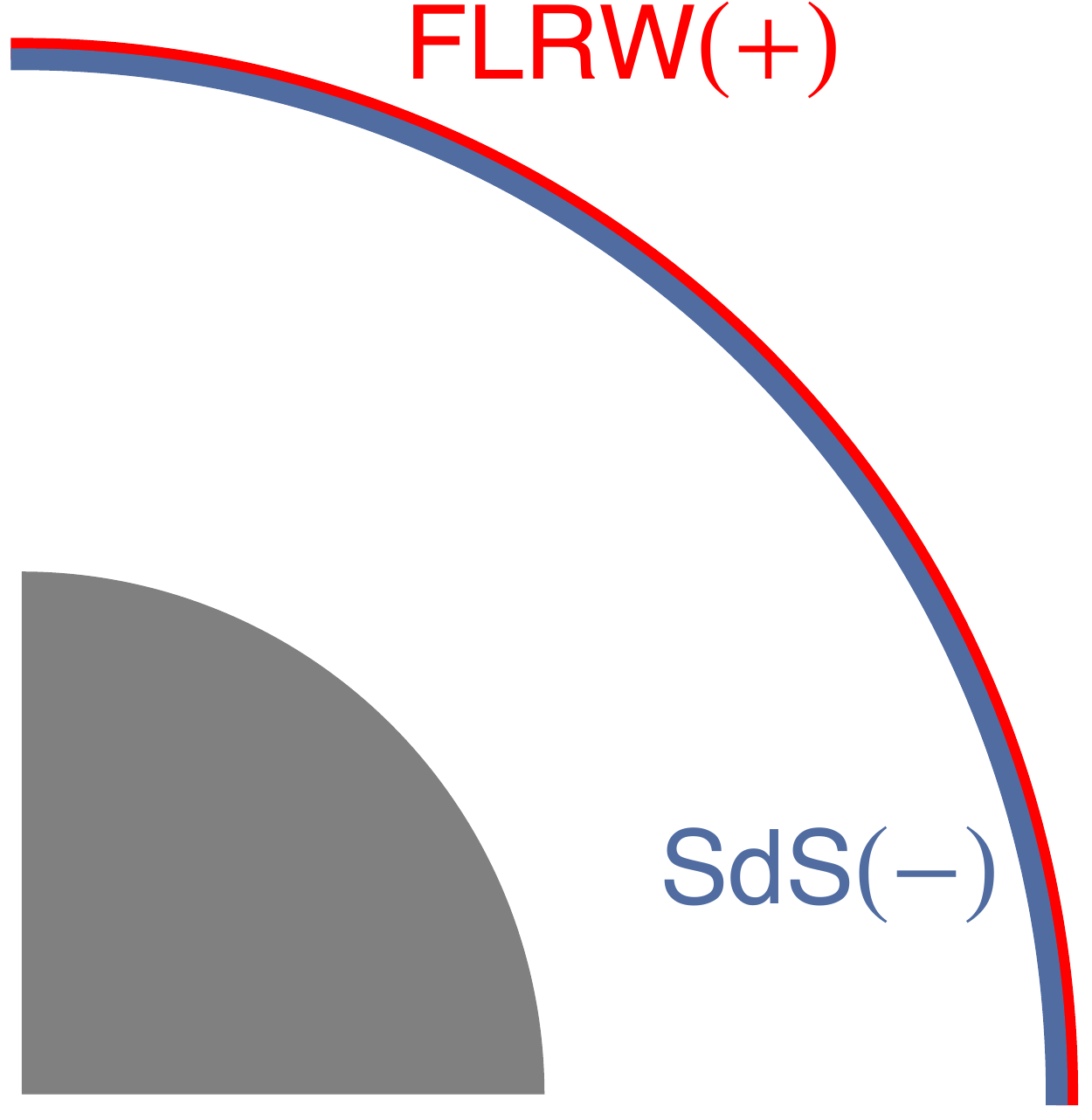}
		\caption{Scheme of quarter of the clump cross section and the matching hypersurface coloured with blue and red stripes representing FLRW side (Red) and SdS side (Blue).\label{FigMH}}
	\end{center}
\end{figure}
The metric induced on the matching hypersurface reads
\begin{equation}
	\diff s_+^2 = -\diff \tau_F^2 + a^2(\tau_F)\Sigma_k^2(\chi_s)\left(\diff\theta^2 + \sin^2\theta\diff\phi^2\right)
\end{equation}
and
\begin{eqnarray}
	\diff s_-^2 &=& -f(r_s)\diff t^2 + h(r_s)\diff r^2_s+ r_s^2\left(\diff\theta^2+\sin^2\theta\diff\phi^2\right)\nonumber\\
		&=&-\diff\tau^2_s + r_s^2\left(\diff\theta^2+\sin^2\theta\diff\phi^2\right)
\end{eqnarray}
where we introduce the proper time of radially receding geodesics $\diff\tau_s$. It clearly reads
\begin{equation}
	\diff\tau^2_s = -f(r_s)\diff t^2 + h(r_s)\diff r^2_s.
\end{equation}
The metric on both sides of the matching hypersurface are identical, i.e. $\diff s_+=\diff s_-$ and the proper-time of the  fundamental observers $\diff \tau_F$ is identical with the proper time of the radially receding geodesics $\diff\tau_s$ implying the condition
\begin{equation}
	r_s(\tau_F)=a(\tau_F)\Sigma_k(\chi_s). \label{junctionCondition}
\end{equation}
\subsection*{Radial Geodesics}
The parameters of the FLRW spacetimes and the parameters of the clump are connected mutually and their connection can be found by the analysis of the radial geodesics of the observers commoving with the clump. It reads
\begin{eqnarray}
	\left(\frac{\diff r_s}{\diff\tau_s}\right)^2&=&\mathcal{E}^2 - 1 + \frac{2M}{r} + \frac{\Lambda}{3}r^2,\label{ur}\\
	\frac{\diff t}{\diff\tau_s}&=&\frac{\mathcal{E}}{f(r)}\label{ut}
\end{eqnarray}
The magnitude of the covariant energy of the radial geodesics must ensure that the radially moving observers with the boundary keep up with the expansion of the FLRW spacetime  that is determined by the Friedmann equation (\ref{Friedmann}) that now reads (applying the  junction condition (\ref{junctionCondition}))
\begin{equation}
	\left(\frac{\diff r_s}{\diff \tau_F}\right)^2=\frac{8\pi}{3}\rho_{m0} r_{s0}^3\frac{1}{r_s} + \frac{\Lambda}{3}r_s^2 - k\,\Sigma_k^2(\chi_s)\label{Friedmann2}.
\end{equation}
Comparing (\ref{ur}) with (\ref{Friedmann2}), one finds out the following identities
\begin{eqnarray}
	\mathcal{E}^2&=&1-k\,\Sigma_k^2(\chi_s),\\
	M&=&\frac{4\pi}{3}\rho_{m0}r_{s0}^3=\frac{H_0^2\Omega_{m0}r_{s0}^3}{2}
\end{eqnarray}
where we have introduced the parameter of matter density  $\Omega_{m0}$ and the Hubble parameter $H_0$ evaluated at our present epoch.

\section{The Temperature Anisotropy of the CMBR }
From the epoch, called "recombination era", the universe becomes transparent and the cosmic radiation moves freely toward us. The effective temperature of this radiation decreases together with the expansion of the Universe $T\sim 1/a$, so we can write
\begin{equation}
	\frac{T_{rec}}{T_{0}}=\frac{a_{0}}{a_{rec}}=(1+z_{rec}).\label{temper1} 
\end{equation}
Now, imagine that we define a region with a commoving radius $\chi_s$. A photon propagating from the LSS enters this region when the scale parameter was $a_i$ and leaves this region when the  scale parameter was $a_o$ and eventually it reaches the observer at our present epoch, having  the scale parameter $a_0$. The formula (\ref{temper1}) can now be written in the form
\begin{equation}
	\frac{T_{rec}}{T_0}=\frac{a_0}{a_o}\frac{a_o}{a_i}\frac{a_i}{a_{rec}}=\frac{r_0}{r_o}\frac{r_o}{r_i}\frac{r_i}{r_{rec}}. \label{temper2} 
\end{equation}
Now, we replace the defined region with the clump. When propagating through the clump a CMBR photon will experience a different frequency shift then in the case of its propagation through the corresponding FLRW. We rewrite (\ref{temper2}) to read
\begin{equation}
	\frac{T_{rec}}{T^c_0}=\frac{r_0}{r_o}(1+z)_c\frac{r_i}{r_{rec}}. \label{temper3} 
\end{equation}
Dividing (\ref{temper3}) by (\ref{temper1}) and introducing $\Delta\equiv r_o/r_i$ we arrive at the  formula
\begin{equation}
	\frac{T_o^c}{T_o}=\frac{\Delta}{(1+z)_c}.\label{temperaniz}
\end{equation}
The frequency shift due to the  clump is determined by the ratio of the frequency  of the  CMBR photon  measured by the  boundary observers when it leaves the clump to  its frequency when it enters the clump, i.e. 
\begin{equation}
	(1+z)_c = \frac{u^\mu k_\mu|_i}{u^\mu k_\mu|_o}.\label{clumpshift}
\end{equation}
Constraining the photon to the motion on the equatorial plane  $\theta=\pi/2$ ,the photon equations of motion in the FLRW and in the SdS regions read
\begin{eqnarray}
	k^{\tau_F}&=&\frac{1}{r_s},\\
	k^\chi&=& \pm\frac{\Sigma_k(\chi_s)}{r_s^2}\sqrt{1-\frac{L^2\,\Sigma_k^2(\chi_s)}{\Sigma_k^2(\chi)}},\\
	k^\phi &=& \frac{L}{r_s^2}\frac{\Sigma_k^2(\chi_s)}{\Sigma_k^2(\chi)}
\end{eqnarray}
and
\begin{eqnarray}
	k^t&=&\frac{E}{f(r)},\label{kt}\\
	k^r&=&\pm\frac{1)}{\sqrt{f(r)h(r)}}\sqrt{E^2-f(r)\frac{L^2}{r^2}},\label{kr}\\
	k^\phi &=&\frac{L}{r^2}.
\end{eqnarray}
Using the formulae (\ref{ur}), (\ref{ut}), (\ref{kr}), and (\ref{kt}) in (\ref{clumpshift}) one arrives at the formula  for the frequency shift in the clump  in the form
\begin{equation}
	(1+z)_c=\frac{f(r_o)}{f(r_i)}\frac{\sqrt{1-k\,\Sigma_k^2(\chi_s)}+\sqrt{(1-f(r_i)L^2/(r_i^2E^2))}}{\sqrt{1-k\,\Sigma_k^2(\chi_s)}-\sqrt{(1-f(r_o)L^2/(r_o^2E^2))}}.\label{zc}
\end{equation}
A CMBR photon is identified with the angular momentum $L$ and the covariant energy $E$. Let $L$ be a free parameter, the  corresponding value of $E$ comes from the fact that at the boundary, both, the fundamental cosmic observer and the  radially moving SdS observer will measure the same value of photon's energy, $u^\alpha k_\alpha$, i.e. ($\vec{u}=u^{\tau_F}\vec{e_{\tau_F}}=u^t\vec{e_t} + u^r\vec{e_r}$)
\begin{eqnarray}
	-k^{\tau_F} &=& u^t k_t + u^r k_r\\
	&\Rightarrow& -\frac{f_o}{r_o}=-E\sqrt{1-k\Sigma_k^2(\chi_s)}+\sqrt{1-f_o-k\Sigma_k^2(\chi_s)}\sqrt{E^2-f_o L^2/r_o^2}\nonumber
\end{eqnarray}
resp.
\begin{equation}
	r_o E = \sqrt{1-k\Sigma_k^2(\chi_s)}+\sqrt{1-L^2}\sqrt{1-f_o-k\Sigma_k^2(\chi_s)}.\label{phenergy}
\end{equation}
In order to determine $\Delta=r_o/r_i$ we just compare the coordinate time interval $\Delta t_c$ it takes the clump to grow from $r_i$ to $r_o$ with the time interval $\Delta t_p$ that elapses between the photon entering the clump for $r_i$ and when it is leaving the clump at $r_o$. One easily finds out the following formulas
\begin{equation}
	\Delta t_c = \int_{r_o/\Delta}^{r_o}\frac{\sqrt{1-k\Sigma_k^2(\chi_s)}\diff r_s}{f(r_s)\sqrt{1-f(r_s)-k\Sigma_k^2(\chi_s)}}\label{dtc}
\end{equation}
and
\begin{eqnarray}
	\Delta t_p &=& \int^{r_o/\Delta}_{r_t}\sqrt{\frac{h(r)}{f(r)}}\frac{\diff r}{\sqrt{1-f(r)L^2/(r^2 E^2)}}\nonumber\\
			 &+& \int^{r_o}_{r_t}\sqrt{\frac{h(r)}{f(r)}}\frac{\diff r}{\sqrt{1-f(r)L^2/(r^2 E^2)}}\label{dtp}
\end{eqnarray}
where is $r_t$ the turning point of the photon geodesics and is the  solution of the equation 
\begin{equation}
	r_t^2 E^2 = f(r)L^2.\label{turpoi}
\end{equation}

\section{Simulations and Results}
We now use the results of our previous sections to determine the effect of the cosmological constant $\Lambda$ and the mass of clump on the ratio $T_0^c/T_0$. The procedure is the following:
\begin{enumerate}
	\item Set up the mass and the density parameters of dark energy  $\Omega_{m0}$ and $\Omega_{\Lambda 0}$ at our present epoch.
	\item Set up the radius of the clump  $r_{s0}$ and the radius of the halo, $R$ (the clump is modelled with a constant density halo) at our present epoch.
	\item The correspondance  between the mass of the  clump  and the cosmological constant are calculated from the  formulas
		\begin{equation}
			M=\frac{H_0^2 \Omega_{m0}r_{s0}^3}{2}
		\end{equation}
		and 
		\begin{equation}
			\Lambda = 3H_0^2\Omega_{\Lambda 0}.
		\end{equation}
	\item Set the value of the free parameter $L\in(0,1)$.
	\item Using the formula (\ref{phenergy}) to determine  the constant of motion of the photon $E$
	\item Solve the equation (\ref{turpoi}) for the turning point $r_t$.
	\item Solve the equation (use (\ref{dtc}) and (\ref{dtp})) 
	\begin{equation}
		\Delta t_c(\Delta)=\Delta t_p(\Delta)
	\end{equation}
	for $\Delta\in [1,\Delta_{max})$.
	\item Determine $(1+z)_c$ from the formula (\ref{zc}).
	\item Determine $T_0^c/T_0$, resp. $\Delta T/T_0\equiv 1 - T_0^c/T_0$ using the formula (\ref{temperaniz}).
\end{enumerate} 
We have prepared the simulation of the temperature anisotropy of the  CMBR to show the effect of the parameter of the  photon  $L$, of the clump's mass, of the clump's model, and the effect of the cosmological constant.

\begin{table}[H]
	\begin{center}
	\caption{Illustrative simulations of the  temperature anisotropy of the CMBR $\Delta T/T_0$ due to the Rees-Sciama effect for a clump with the mass $\sim 10^{19}M_{\odot}$.}\label{Table1}
	\begin{tabular}{|c|c|c|c|c|c|c|c|}
		\hline
		Model & $R[\mathrm{Mpc}]$ & $r_{s0}[\mathrm{Mpc}]$ & $r_o[\mathrm{Mpc}]$ & $\Omega_{m0}$ & $\Omega_{\Lambda 0}$ & $L$ & $\Delta T/T_0$\\
		\hline
		BH & - & $500.0$ & $400.0$ & $0.25$ & $0.75$ & $0.5$ & $-2.7\times 10^{-4}$ \\
		BH & - & $500.0$ & $300.0$ & $0.25$ & $0.75$ & $0.5$ & $-3.5\times 10^{-4}$\\
		BH & - & $500.0$ & $200.0$ & $0.25$ & $0.75$ & $0.5$ & $-5.9\times 10^{-4}$\\
		H & $100.0$ & $500.0$ & $400.0$ & $0.25$ & $0.75$ & $0.5$ & $-2.5\times 10^{-4}$\\
		H & $100.0$ & $500.0$ & $300.0$ & $0.25$ & $0.75$ & $0.5$ & $-3.5\times 10^{-4}$\\
		H & $100.0$ & $500.0$ & $200.0$ & $0.25$ & $0.75$ & $0.5$ & $-1.9\times 10^{-3}$\\
		\hline
		BH & - & $500.0$ & $400.0$ & $0.75$ & $0.25$ & $0.5$ & $-1.1\times 10^{-3}$\\
		BH & - & $500.0$ & $300.0$ & $0.75$ & $0.25$ & $0.5$ & $-1.7\times 10^{-3}$\\
		BH & - & $500.0$ & $200.0$ & $0.75$ & $0.25$ & $0.5$ & $-3.5\times 10^{-3}$\\
		H & $100.0$ & $500.0$ & $400.0$ & $0.75$ & $0.25$ & $0.5$ & $-1.1\times 10^{-3}$\\
		H & $100.0$ & $500.0$ & $300.0$ & $0.75$ & $0.25$ & $0.5$ & $-1.7\times 10^{-3}$\\
		H & $100.0$ & $500.0$ & $200.0$ & $0.75$ & $0.25$ & $0.5$ & $-1.1\times 10^{-2}$\\
		\hline
	\end{tabular}
	\end{center}
\end{table}

\begin{table}[H]
\begin{center}
	\caption{Illustrative simulations of the temperature anisotropy of the CMBR $\Delta T/T_0$ due to the  Rees-Sciama effect for clump with the mass $\sim 10^{18}M_{\odot}$.}\label{Table2}
	\begin{tabular}{|c|c|c|c|c|c|c|c|}
		\hline
		Model & $R[\mathrm{Mpc}]$ & $r_{s0}[\mathrm{Mpc}]$ & $r_o[\mathrm{Mpc}]$ & $\Omega_{m0}$ & $\Omega_{\Lambda 0}$ & $L$ & $\Delta T/T_0$\\
		\hline
		BH & - & $250.0$ & $200.0$ & $0.25$ & $0.75$ & $0.5$ & $-3.3\times 10^{-5}$\\
		BH & - & $250.0$ & $180.0$ & $0.25$ & $0.75$ & $0.5$ & $-3.5\times 10^{-5}$\\
		BH & - & $250.0$ & $150.0$ & $0.25$ & $0.75$ & $0.5$ & $-4.1\times 10^{-5}$\\
		H & $100.0$ & $250.0$ & $200.0$ & $0.25$ & $0.75$ & $0.5$ & $3.9 \times 10^{-4}$ \\
		H & $100.0$ & $250.0$ & $180.0$ & $0.25$ & $0.75$ & $0.5$ & $4.8\times 10^{-4}$\\
		H & $100.0$ & $250.0$ & $150.0$ & $0.25$ & $0.75$ & $0.5$ & $-3.3\times 10^{-4}$\\
		\hline
		BH & - & $250.0$ & $200.0$ & $0.75$ & $0.25$ & $0.5$ & $-4.1\times 10^{-4}$\\
		BH & - & $250.0$ & $180.0$ & $0.75$ & $0.25$ & $0.5$ & $-1.1\times 10^{-3}$\\
		BH & - & $250.0$ & $150.0$ & $0.75$ & $0.25$ & $0.5$ & $-2.4\times 10^{-3}$\\
		H & $100.0$ & $250.0$ & $200.0$ & $0.75$ & $0.25$ & $0.5$ & $-4.1\times 10^{-4}$\\
		H & $100.0$ & $250.0$ & $180.0$ & $0.75$ & $0.25$ & $0.5$ & $-1.1\times 10^{-3}$\\
		H & $100.0$ & $250.0$ & $150.0$ & $0.75$ & $0.25$ & $0.5$ & $-2.4\times 10^{-3}$\\
		\hline
	\end{tabular}
	\end{center}
\end{table}

\begin{figure}[H]
	\begin{center}
	\begin{tabular}{c}
		\includegraphics[scale=0.7]{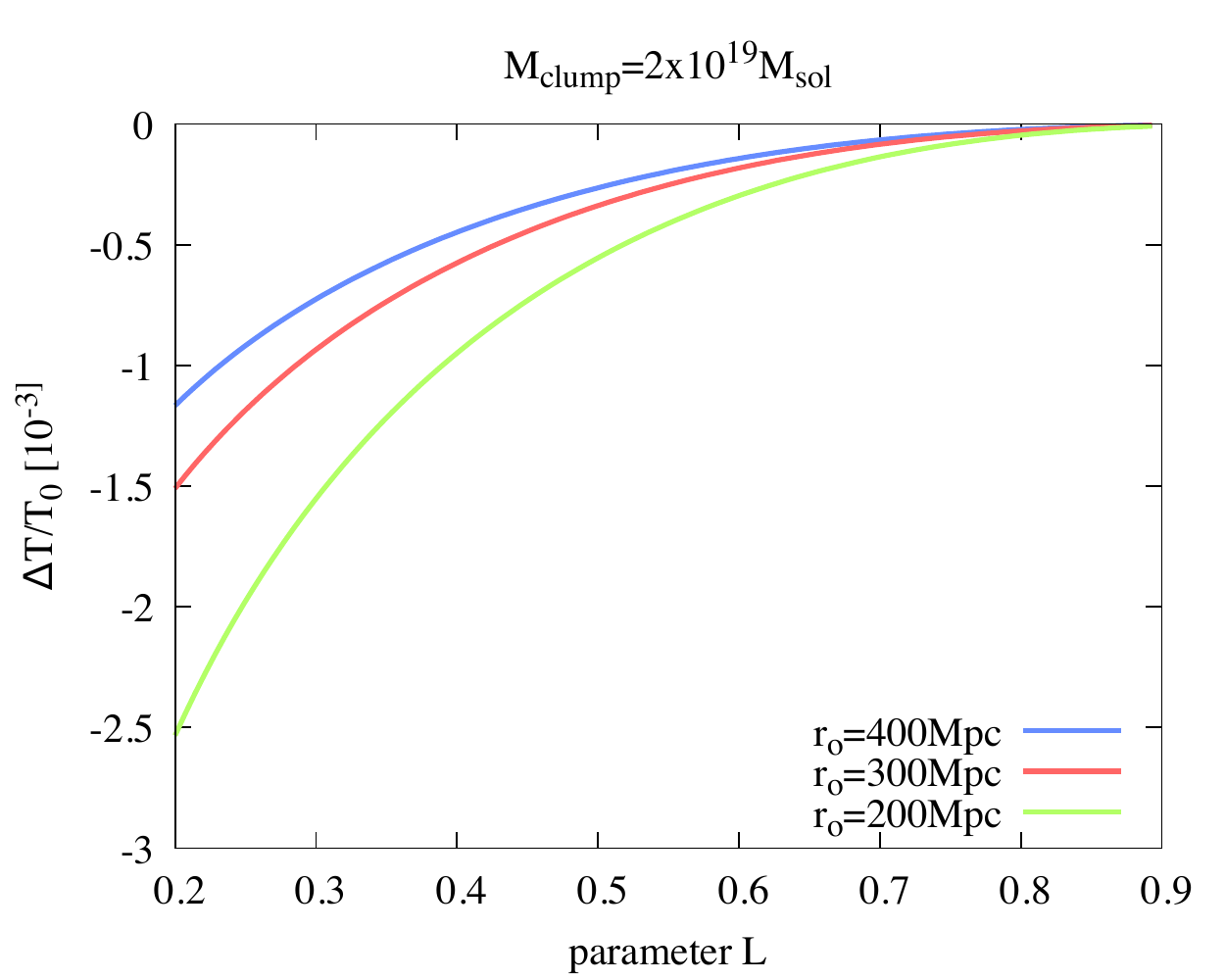}\\
		\includegraphics[scale=0.7]{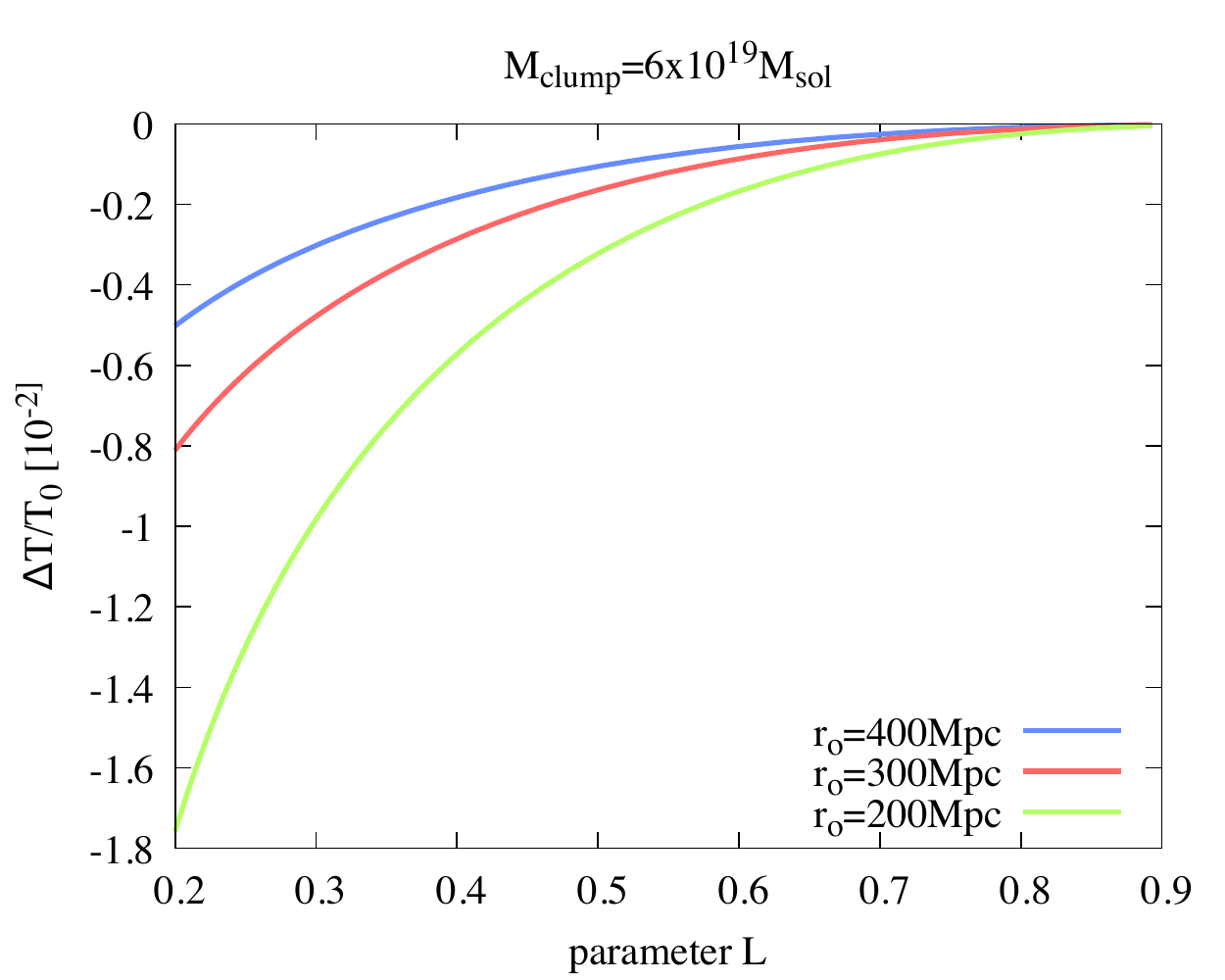}
	\end{tabular}
		\caption{Illustrative simulations of the temperature anisotropy of the CMBR due to \emph{black-hole} clump,  with a mass $\sim 10^{19}M_{\odot}$, as a function of the parameter $L$. The spatial geometry of the universe is flat with the  density parameters: left panel - $\Omega_{m0}=0.25$, $\Omega_{\Lambda 0}=0.75$, right panel - $\Omega_{m0}=0.75$, $\Omega_{\Lambda 0}=0.25$.\label{Fig3}}
		\end{center}
\end{figure}

\begin{figure}[H]
	\begin{center}
	\begin{tabular}{c}
		\includegraphics[scale=0.7]{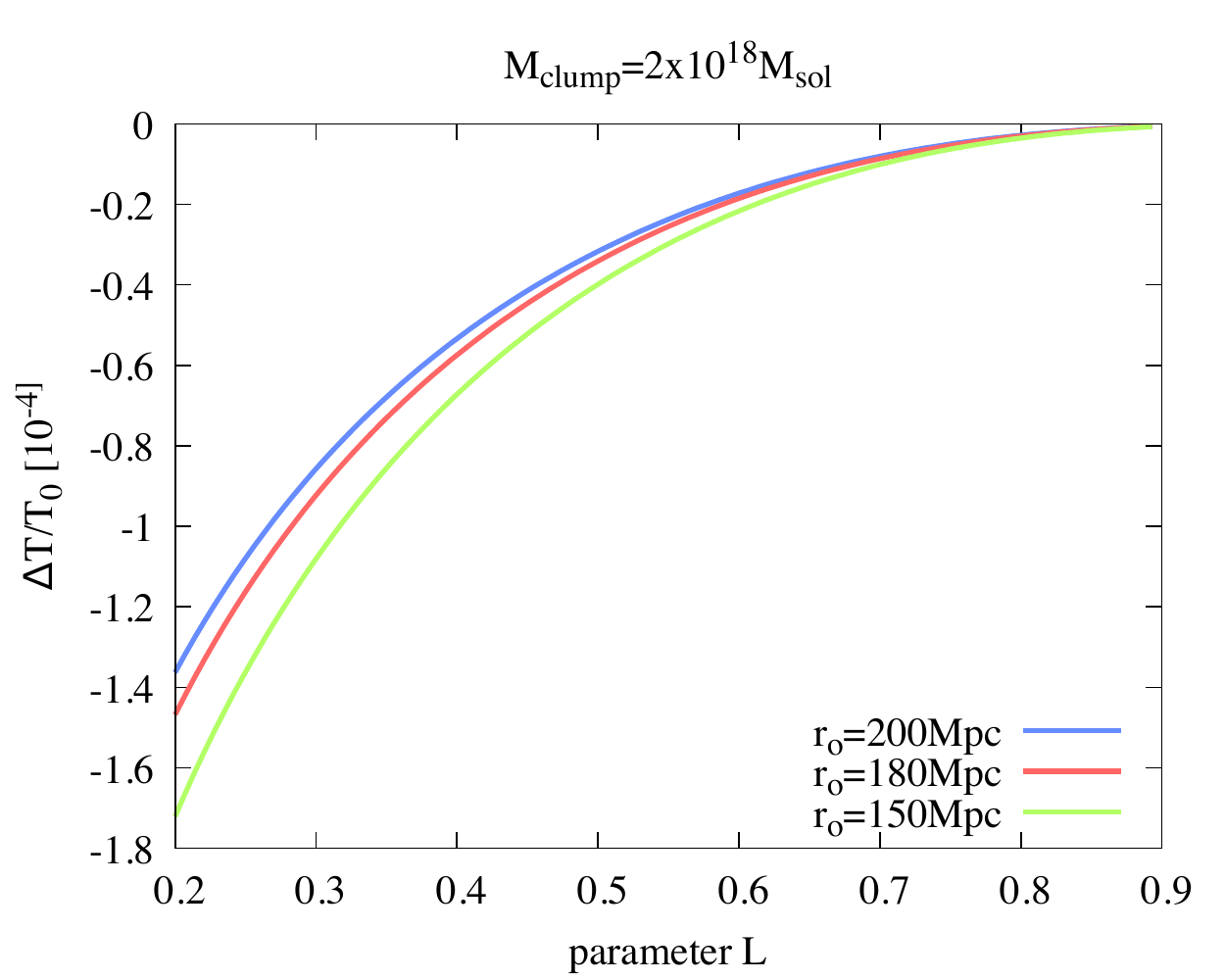}\\
		\includegraphics[scale=0.7]{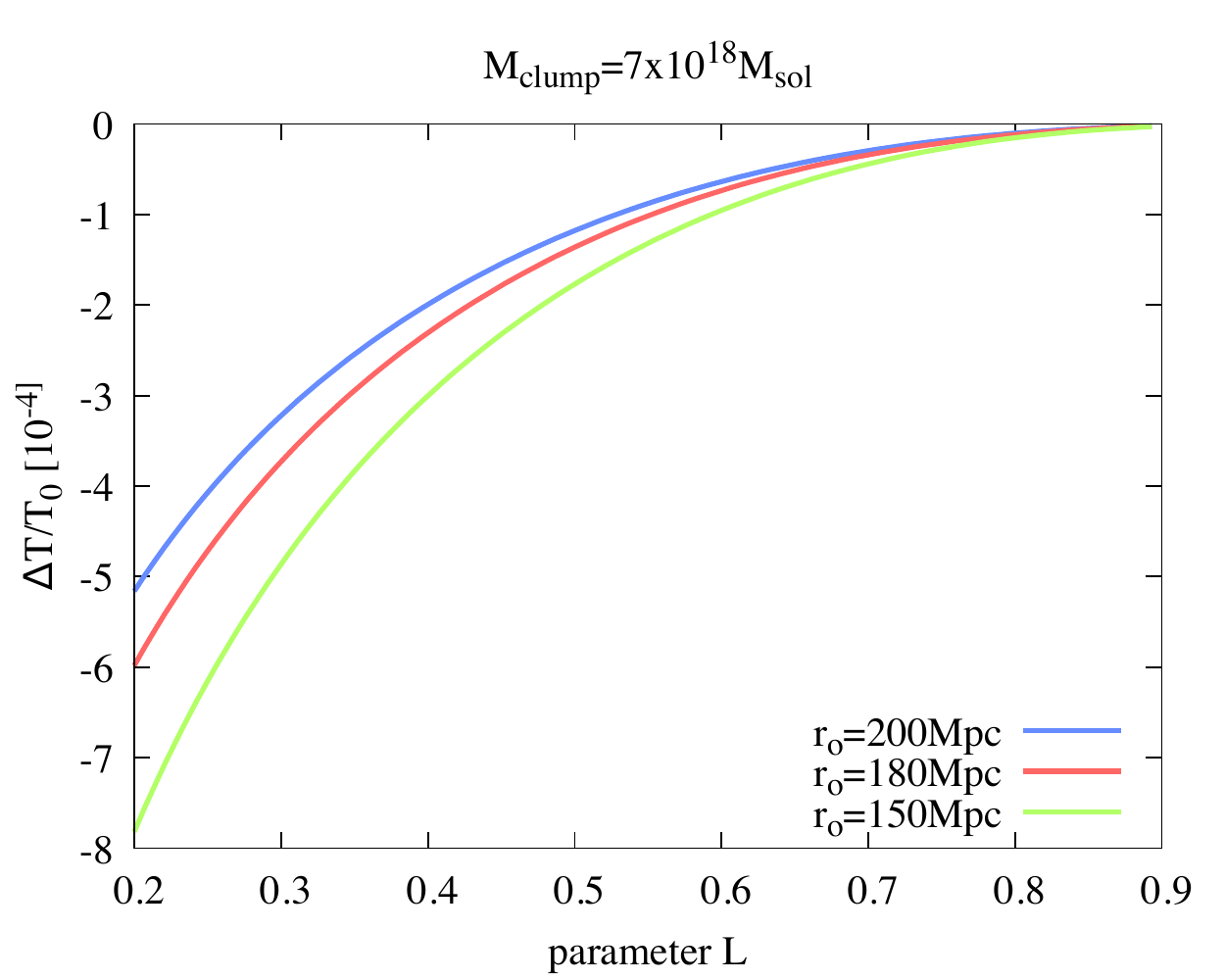}
	\end{tabular}
		\caption{Illustrative simulations of the temperature anisotropy of the CMBR due to \emph{black-hole} clump, with a mass $\sim 10^{18}M_{\odot}$, as a function of the parameter $L$. The spatial geometry of the universe is flat with the density parameters: left panel - $\Omega_{m0}=0.25$, $\Omega_{\Lambda 0}=0.75$, right panel - $\Omega_{m0}=0.75$, $\Omega_{\Lambda 0}=0.25$.\label{Fig4}}
	\end{center}
\end{figure}

\begin{figure}[H]
	\begin{center}
	\begin{tabular}{c}
		\includegraphics[scale=0.7]{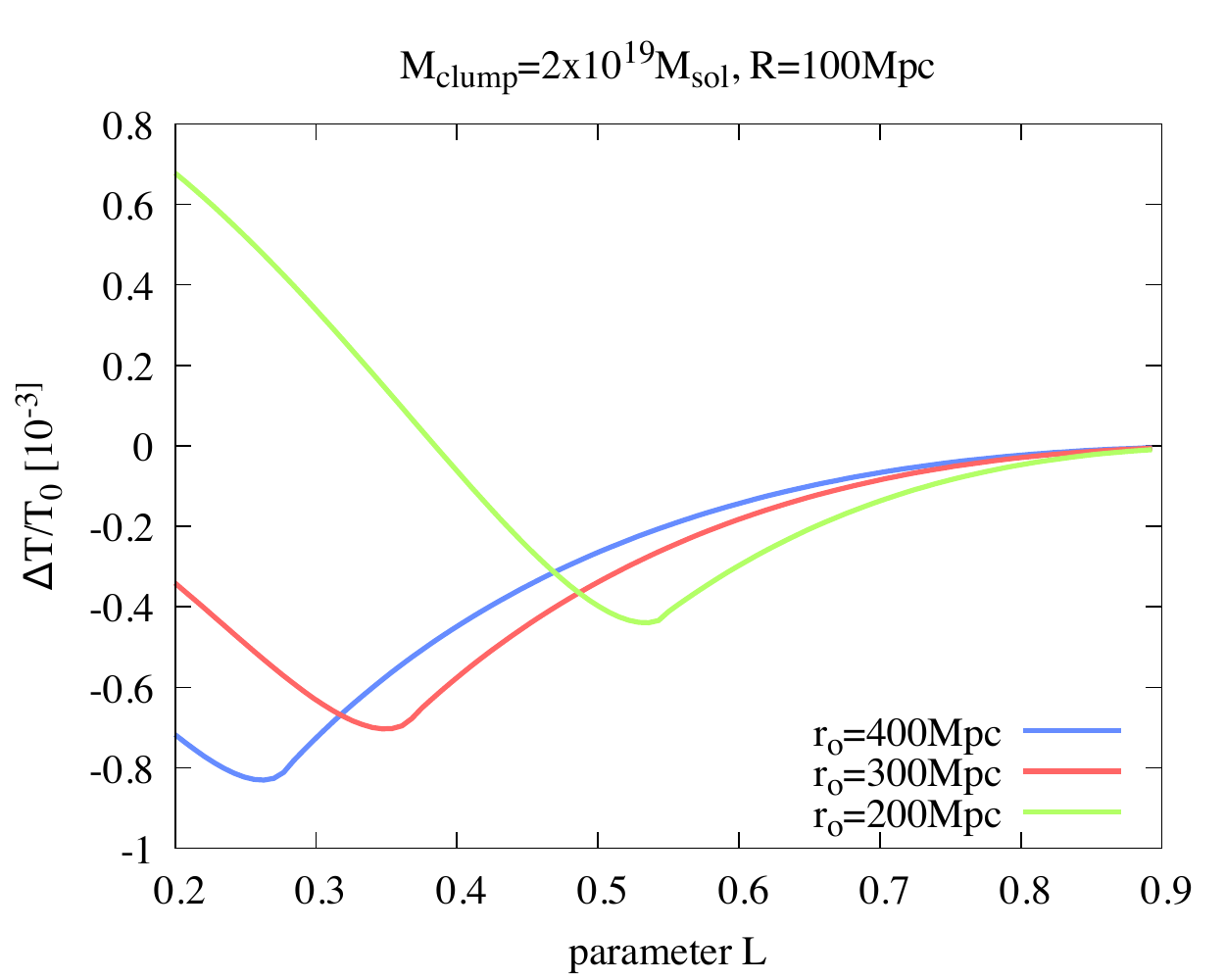}\\
		\includegraphics[scale=0.7]{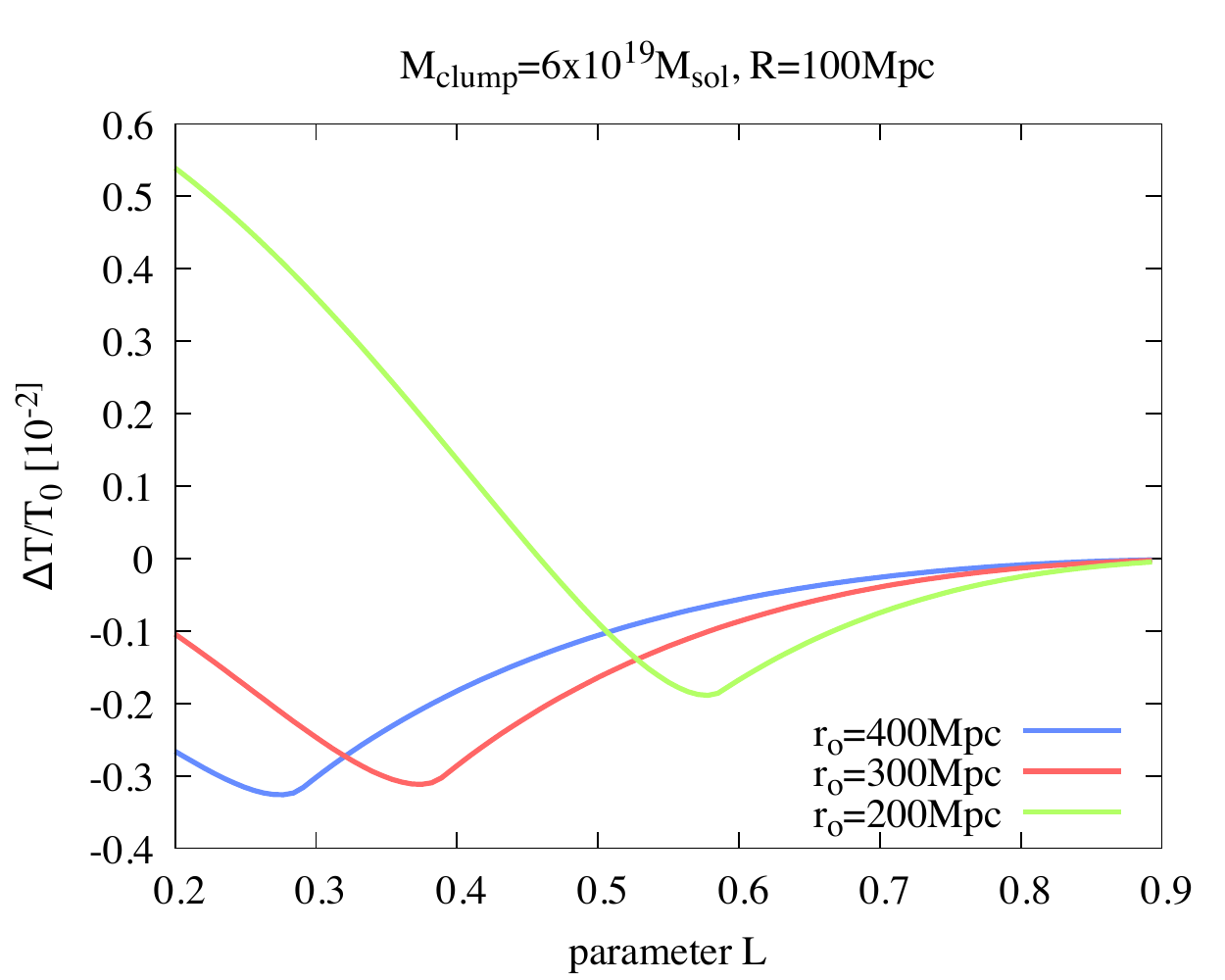}
	\end{tabular}
		\caption{Illustrative simulations of the temperature anisotropy of the  CMBR  due to \emph{constant density halo} clump,  with the mass $\sim 10^{19}M_{\odot}$, as a function of the parameter $L$. The spatial geometry of the universe is flat with the density parameters: left panel - $\Omega_{m0}=0.25$, $\Omega_{\Lambda 0}=0.75$, right panel - $\Omega_{m0}=0.75$, $\Omega_{\Lambda 0}=0.25$.\label{Fig5}}
	\end{center}
\end{figure}

\begin{figure}[H]
	\begin{center}
	\begin{tabular}{c}
		\includegraphics[scale=0.7]{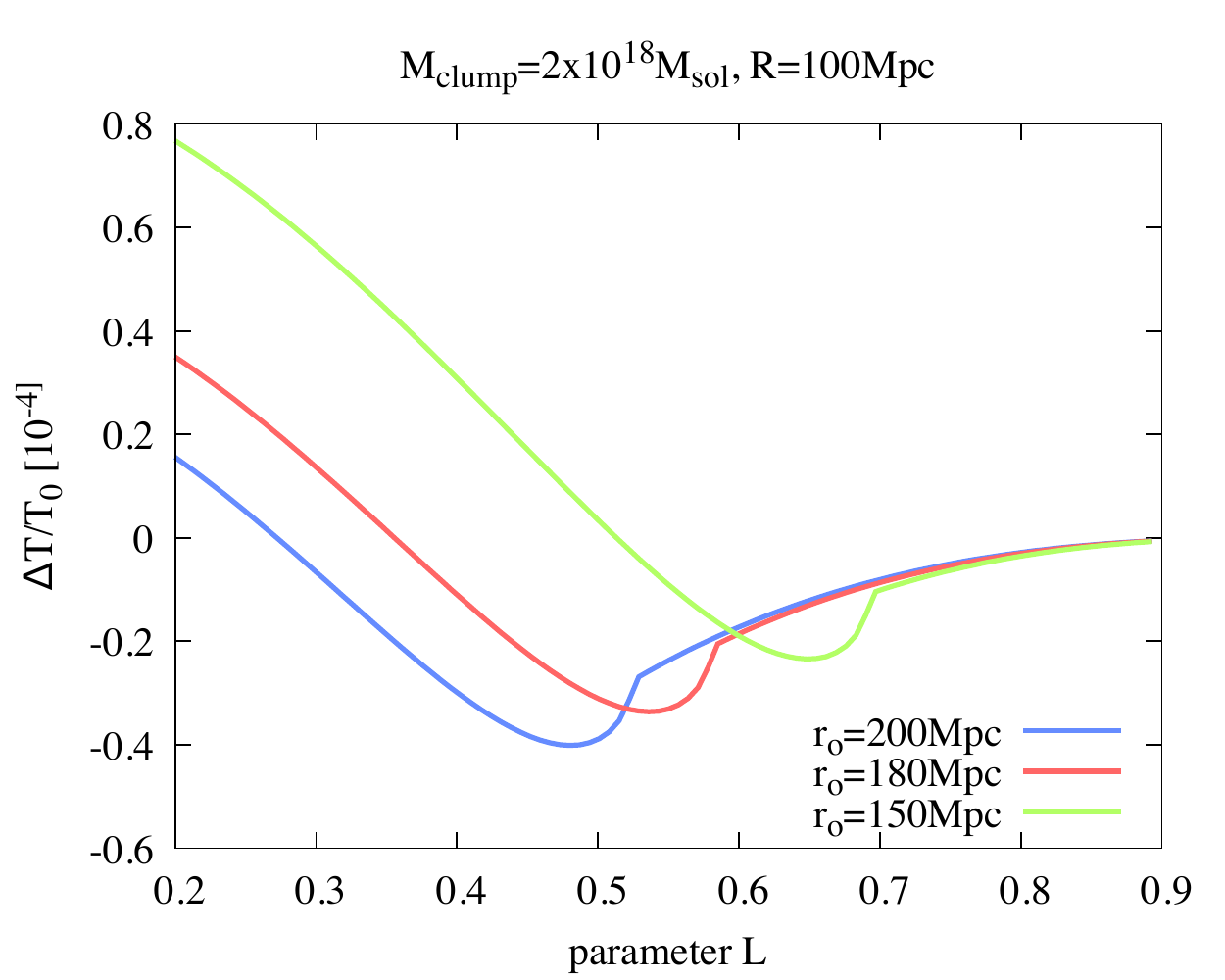}\\
		\includegraphics[scale=0.7]{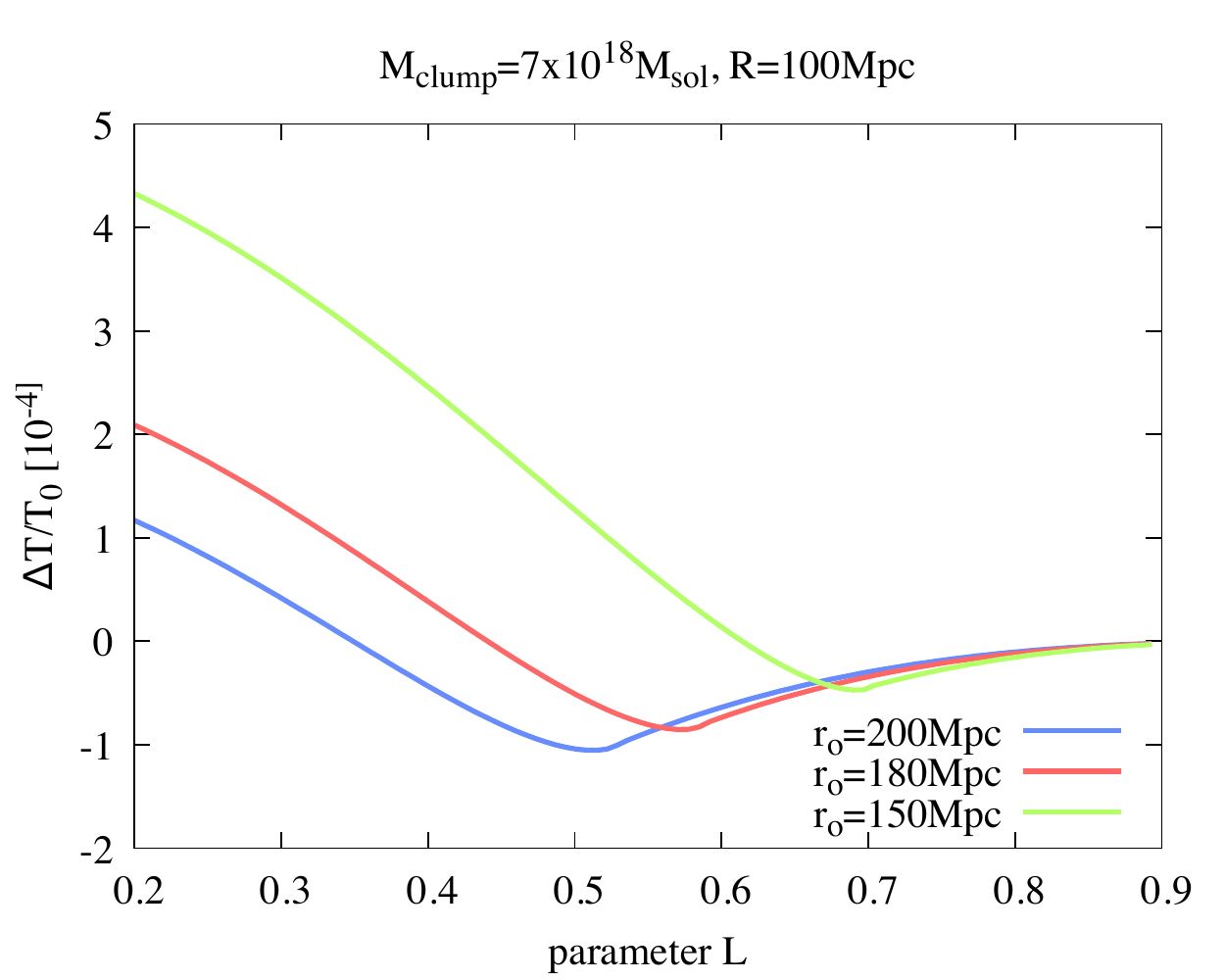}
	\end{tabular}
		\caption{Illustrative simulations of the temperature anisotropy of the CMBR due to  the \emph{constant density halo} clump,  with a mass $\sim 10^{18}M_{\odot}$, as a function of the parameter $L$. The spatial geometry of the universe is flat with the density parameters: left panel - $\Omega_{m0}=0.25$, $\Omega_{\Lambda 0}=0.75$, right panel - $\Omega_{m0}=0.75$, $\Omega_{\Lambda 0}=0.25$.\label{Fig6}}
	\end{center}
\end{figure}

\begin{figure}[H]
	\begin{center}
	\begin{tabular}{c}
		\includegraphics[scale=0.7]{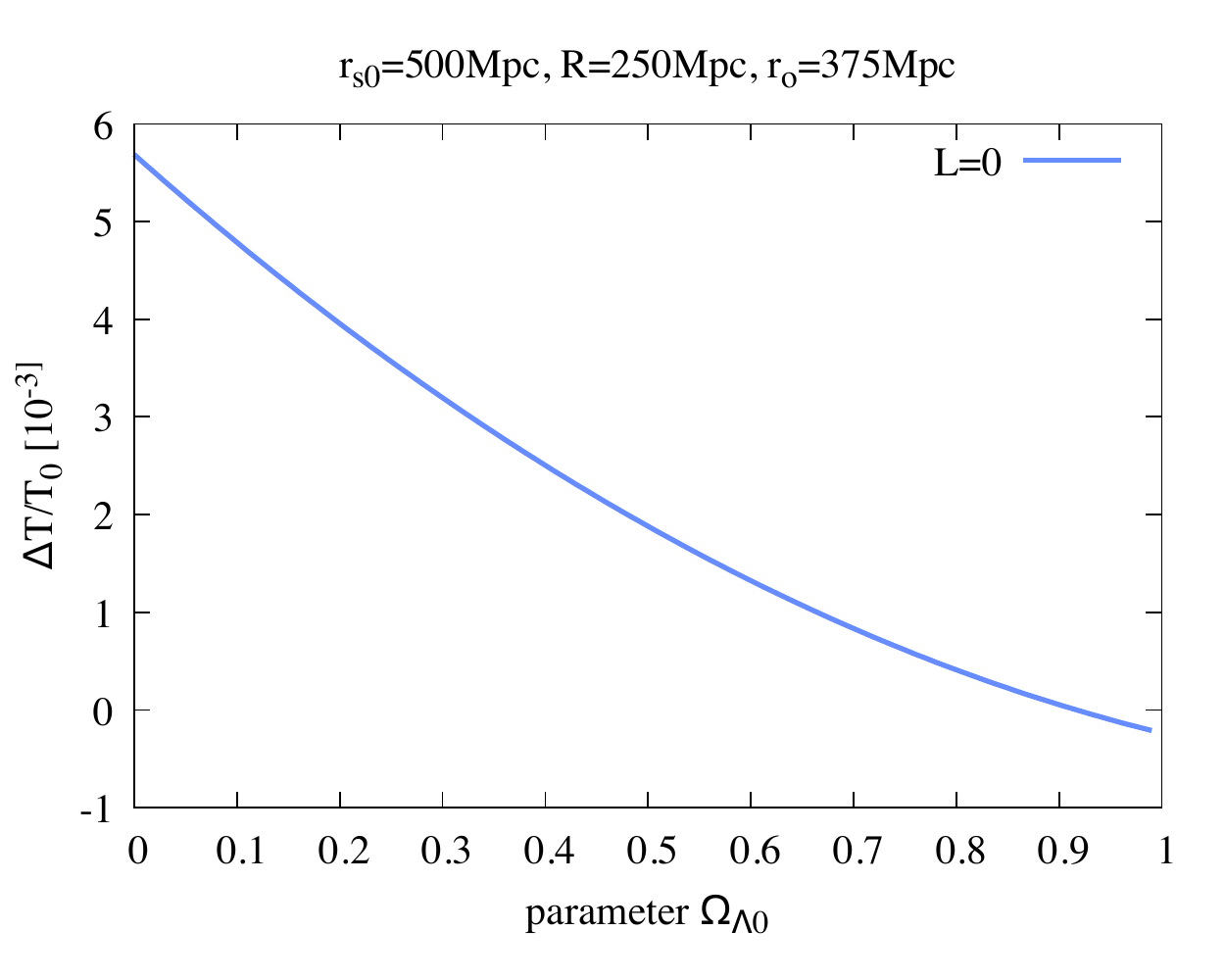}\\
		\includegraphics[scale=0.7]{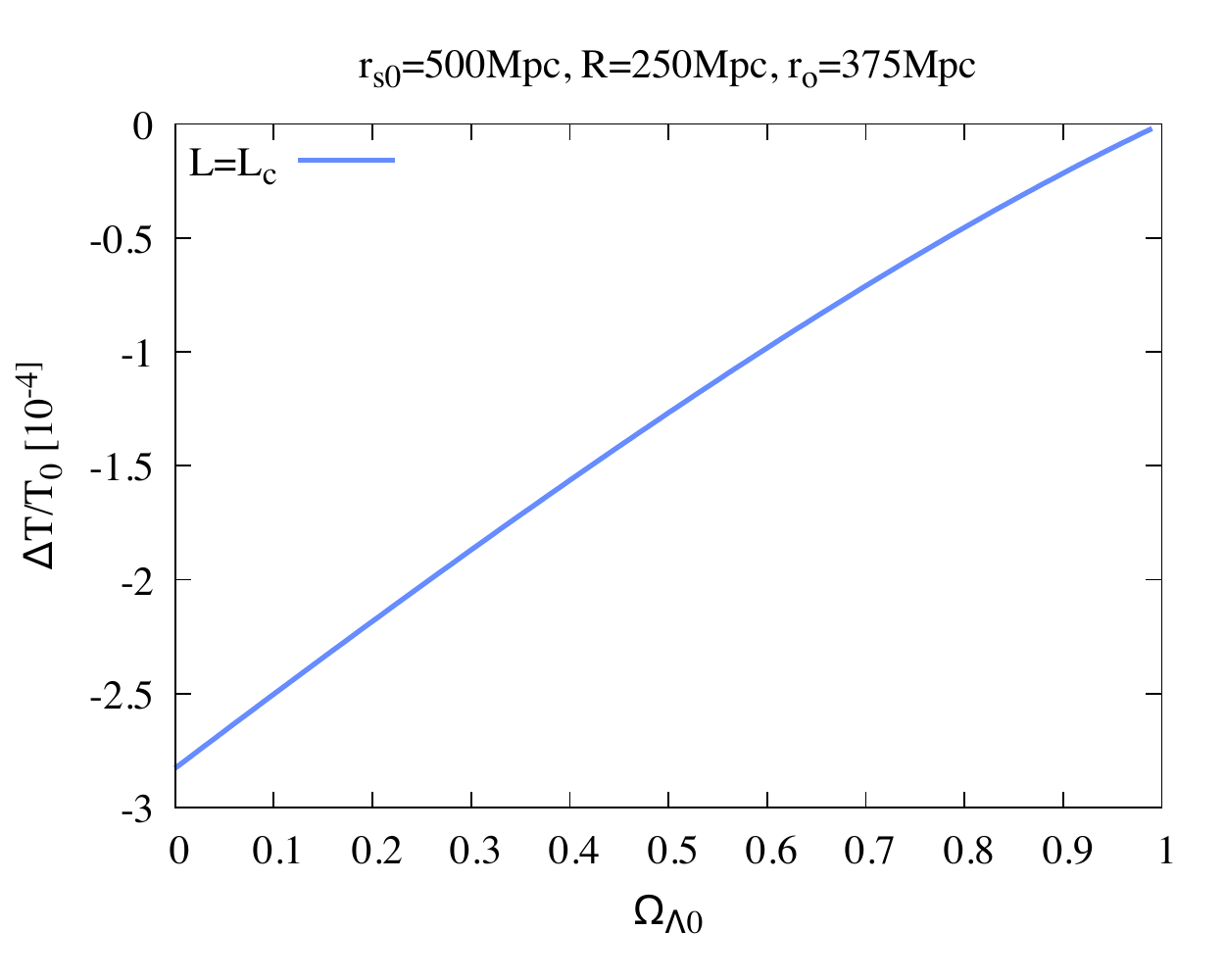}
	\end{tabular}
	\caption{The effect of the cosmological constant $\Lambda$  on the anisotropy of the CMBR. The Cosmological constant is here represented by a corresponding density parameter $\Omega_{\Lambda 0}$. Since we consider the flat universe, it is paired with the baryonic mass density $\Omega_{m0}$ via the formula $1=\Omega_{m0} + \Omega_{\Lambda 0}$ (the radiation energy density is considered negligible). The clump is represented by a halo with constant density with a  mass of about $10^{19}M_\odot$.\label{Fig7}}
	\end{center}
\end{figure}

\begin{figure}[H]
	\begin{center}
	\begin{tabular}{c}
		\includegraphics[scale=0.7]{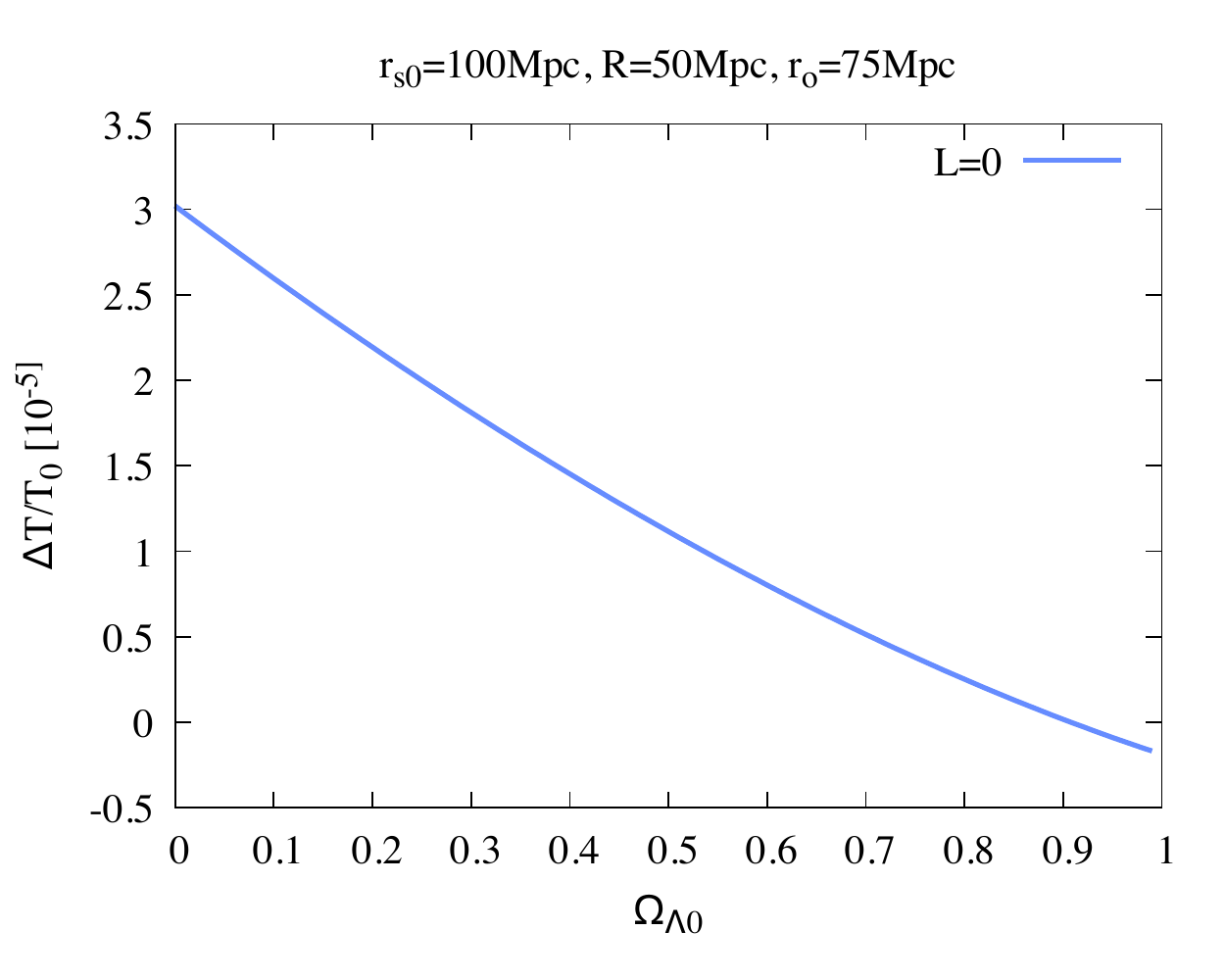}\\
		\includegraphics[scale=0.7]{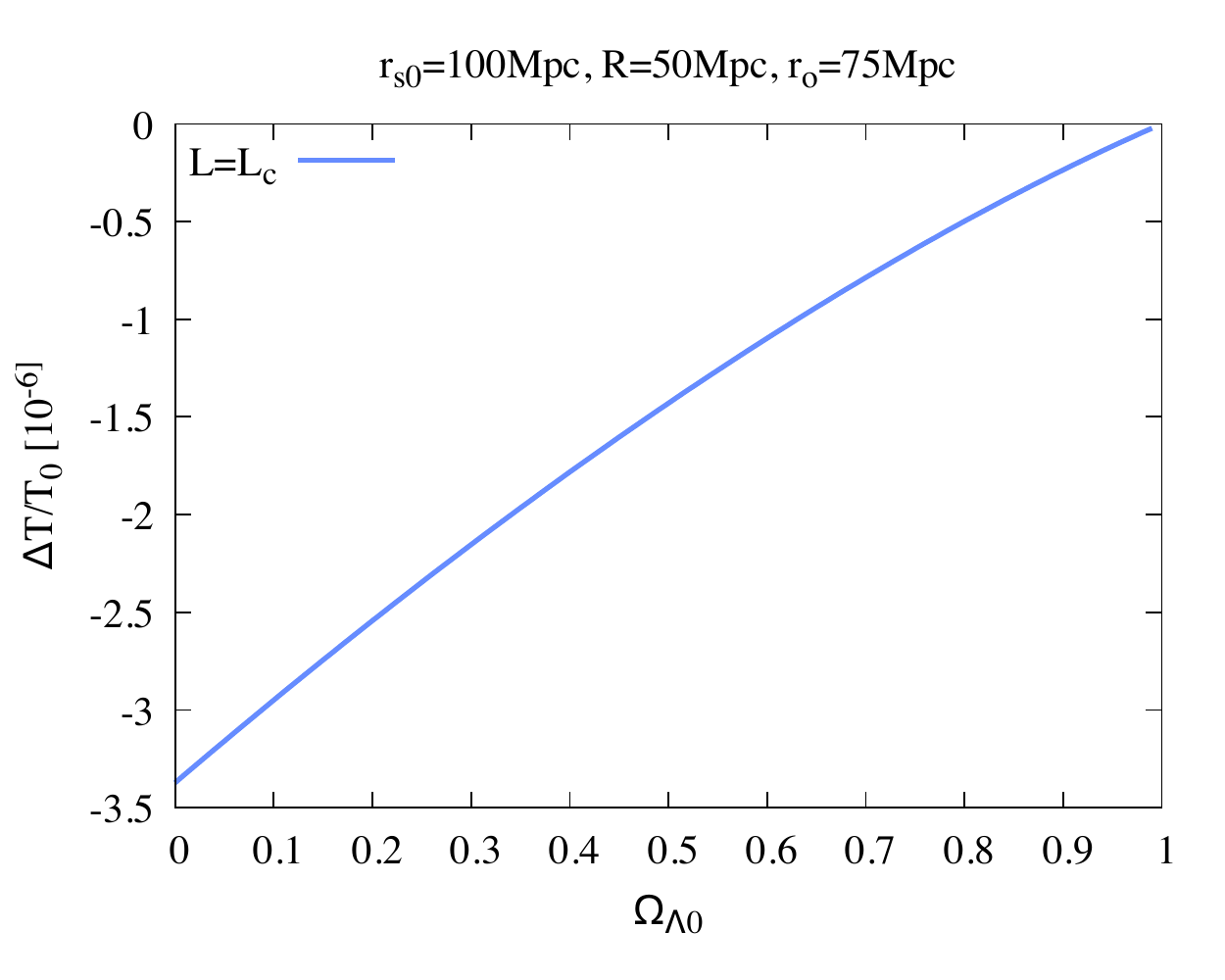}
	\end{tabular}
	\caption{The effect of the cosmological constant $\Lambda$ on the anisotropy of the CMBR. The  Cosmological constant is here represented by a corresponding density parameter $\Omega_{\Lambda 0}$. Since we consider the flat universe it is paired with the baryonic mass density $\Omega_{m0}$ via formula $1=\Omega_{m0} + \Omega_{\Lambda 0}$ (radiation energy density is considered negligible). The clump is represented by a halo with constant density with a  mass of about $10^{18}M_\odot$. \label{Fig8}}
	\end{center}
\end{figure}

\begin{figure}[H]
	\begin{center}
	\begin{tabular}{c}
		\includegraphics[scale=0.7]{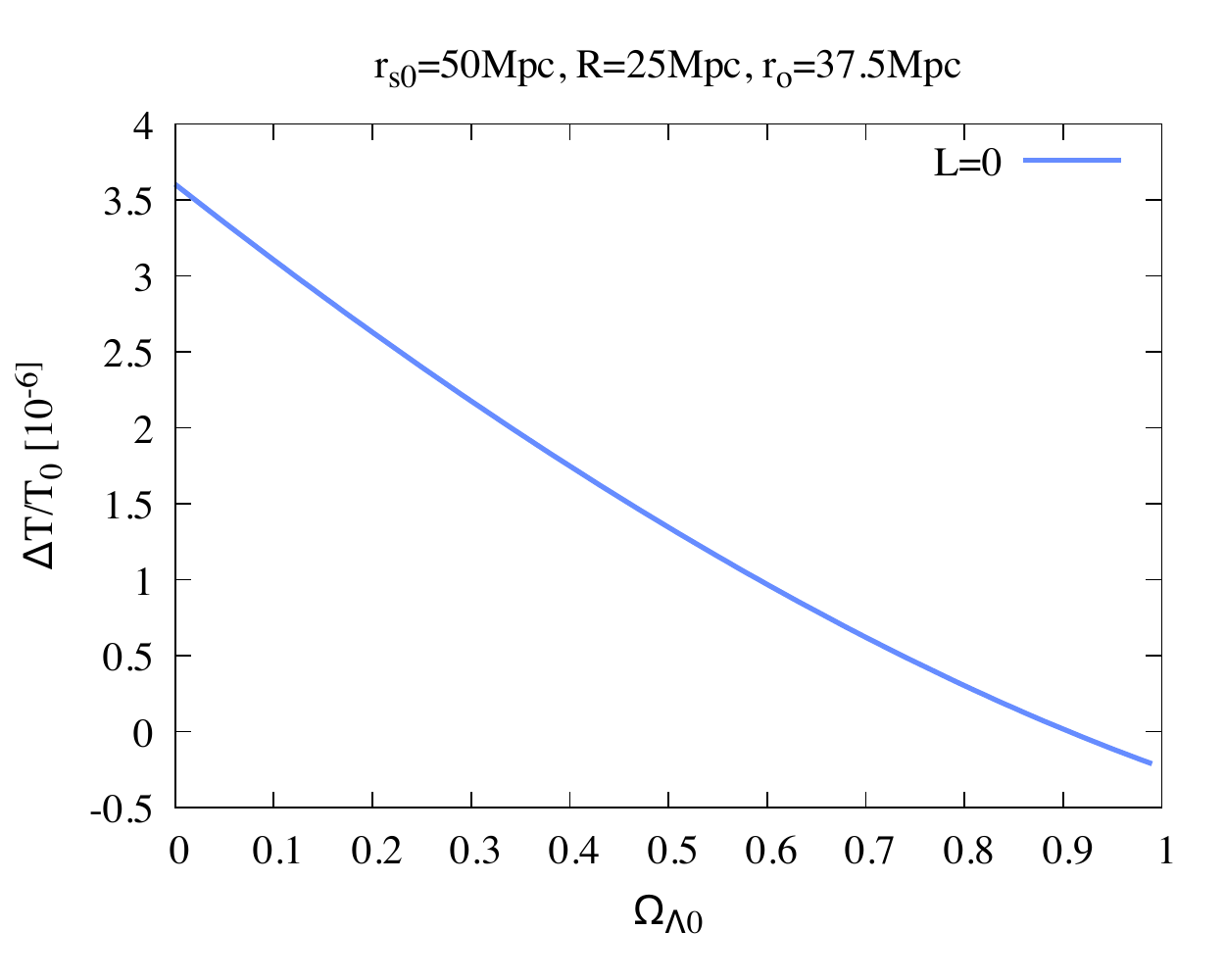}\\
		\includegraphics[scale=0.7]{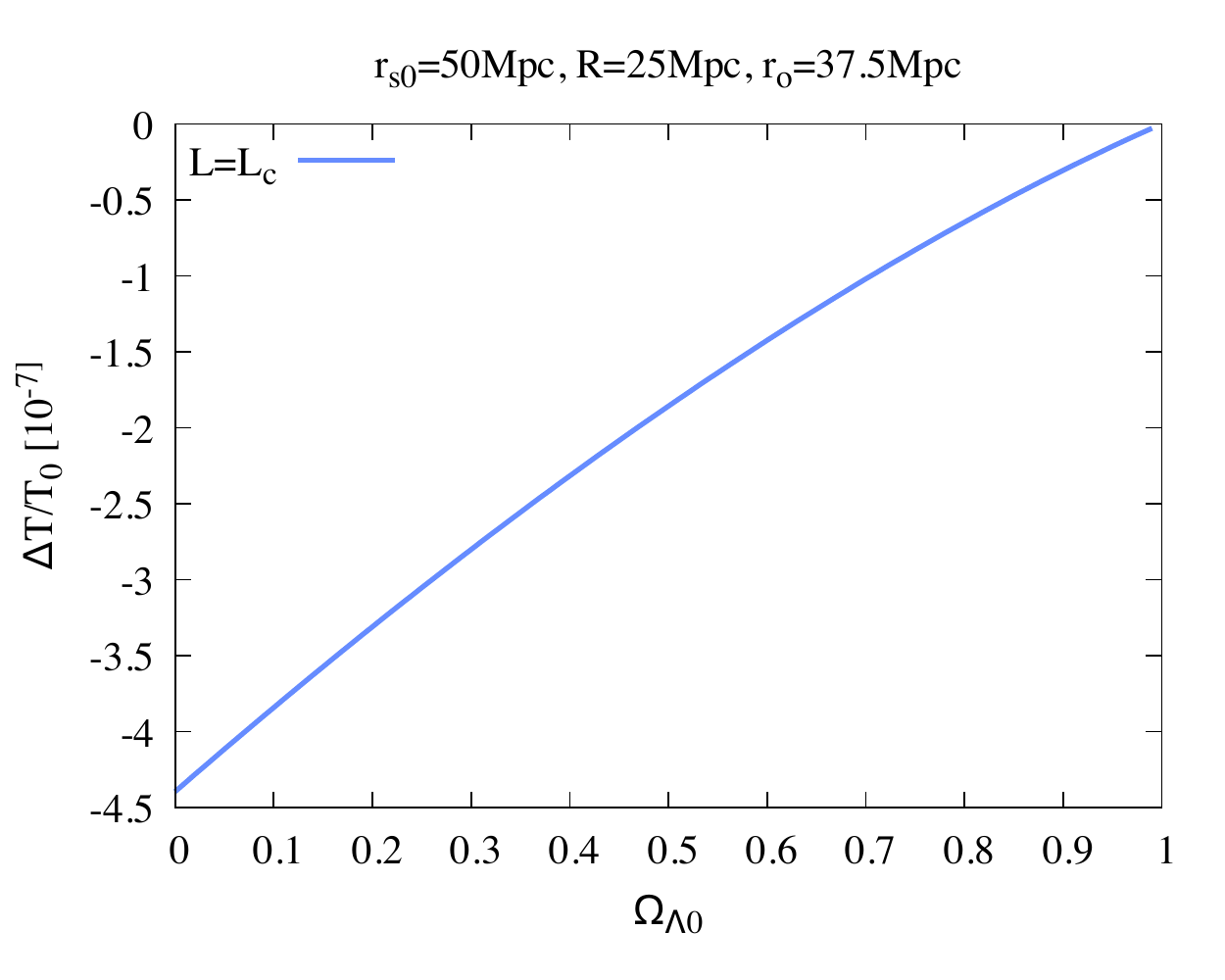}
	\end{tabular}
	\caption{The effect of the cosmological constant $\Lambda$  on the anisotropy of the CMBR. The Cosmological constant is here represented by a corresponding density parameter $\Omega_{\Lambda 0}$. Since we consider the flat universe it is paired with the baryonic mass density $\Omega_{m0}$ via formula $1=\Omega_{m0} + \Omega_{\Lambda 0}$ (radiation energy density is considered negligible). The clump is represented by a halo with constant density with a  mass of about $10^{16}M_\odot$.\label{Fig9}}
	\end{center}
\end{figure}


\section{Discussion and Conclusions}
According to the observations \cite{Bennett_etal:2013,Planck:2018:AA} we considered in our simulations positive values of the cosmological constant and that the spatial geometry of the FLRW is flat ($k=0$). For the understanding of the results it is worth to introduce the quantity $\Delta_{RW}$ that represents the  cosmological redshift of the CMBR photon crossing the region of a commoving radius $\chi_S$. We define the  time-delay effect due to the clump as the ratio $\Delta/\Delta_{RW}$ and  the redshift effect as $(1+z)_c/\Delta_{RW}$. When the time-delay (redshift) effects dominate the temperature anisotropy then $\Delta T/T_0<0$ ($\Delta T/T_0>0$). When there is $\Delta T/T_0=0$ both effects cancel each other.

We have prepared three simulation setups. First, the simulations of the Rees-Sciama effect in the case of a clump with a mass $\sim 10^{19}\mathrm{M}_\odot$ (Table \ref{Table1}) and a clump with a mass $\sim 10^{18}\mathrm{M}_\odot$ (Table \ref{Table2}). Two kinds of clumps are here considered, the black hole (BH) and the constant density halo (H). There are two key regimes of the simulations. First the dark energy is dominant, as follows from the recent observations of WMAP and Planck. We clearly see that the absolute value of anisotropies of the temperature of the CMBR  is $10$-times smaller than in the opposite case when the dark and baryonic matter are dominant. This is expected due to a stronger gravitational redshift in the second regime. Further, when the mass is $10$-times smaller than the temperature anisotropies are, approximately, $10$-times smaller too.

 Secondly, we have constructed the CMBR temperature anisotropy profile with respect to the photon angular momentum parameter $L$, Figures \ref{Fig3} - \ref{Fig6}. In the corresponding plots three curves are plotted for three representative values of the radius $r_o$ (radius of the clump in the moment when the  CMBR photon emerge from the clump). When the clump is formed out of the black-hole, Figs. \ref{Fig3} and \ref{Fig4}, the larger is the value of $r_o$ the smaller is the absolute value of the  temperature anisotropy and the larger is the value of $L$ the smaller is the temperature anisotropy. Notice that there is $\Delta T/T_0<0$ which means that the time delay effect dominates over the redshift effect. When the clump model is a halo with  constant density, Figs. \ref{Fig5} and \ref{Fig6}, then the profile of $\Delta T/T_0$ depends on the geodesics of the CMBR photon. If the geodesics does not cross the halo, i.e. $r_t>R$ then, clearly, the behaviour of $\Delta T/T_0 (L)$ is the same as in the case of black-hole clump. In case of $r_t<R$ the geodesics crosses the halo and the behaviour of the temperature anisotropy is the opposite to the black hole clump case. One can also observe that for each fixed value $r_o$ there exists $L$ where the time-delay and redshift effects cancel, it is a point where the dominance between time-delay effects and redshift effects exchange.
 
Third, the simulations are focused on the effect of $\Omega_{\Lambda 0}$ on $\Delta T/T_0$ for two representative values of $L=0$, $L=L_c$ (Figs. \ref{Fig7} - \ref{Fig9}) where $L_c$ is the value of $L$ for $r_t=R$. In all four cases the behaviour of $\Delta T/T_0$ is monotonic. For $L=0$ then $\Delta T/T_0$ decreases with increasing value of $\Omega_{\Lambda 0}$. There exists $\Omega_{\Lambda 0(c)}$ where $\Delta T/T_0=0$. For $\Omega_{\Lambda 0}<\Omega_{\Lambda 0(c)}$ the redshift effect dominates while for $\Omega_{\Lambda 0}>\Omega_{\Lambda 0(c)}$ the time-delay effect is dominant.  For $L=L_c$ the absolute value of $\Delta T/T_0$ decreases and for the whole interval of $\Omega_{\Lambda 0}$ the time-delay effect dominates.

We can conclude that the more dark energy  dominates, the the smaller is the temperature anisotropy caused by the Rees-Sciama effect and also the more massive is the clump the larger is the temperature anisotropy.

\section*{Acknowledgement}
 This work was supported by the Student Grant Foundation of the Silesian University in Opava, Grant No. SGF/1/2021, which was realised within the EU OPSRE project entitled "Improving the quality of the internal grant scheme of the Silesian University in Opava", reg. number: $CZ.02.2.69/0.0/0.0/19\_073/0016951$\begin{dedication}
In the  memory of my grandfather Norbert Boj
\end{dedication}

\end{document}